\documentclass[aps,twocolumn,showpacs]{revtex4}

\usepackage{graphicx,amssymb,epsfig,feynmp}
\newcommand{\be}{\begin{equation}}
\newcommand{\ee}{\end{equation}}
\newcommand{\bea}{\begin{eqnarray}}
\newcommand{\eea}{\end{eqnarray}}

\newcommand{\bp}{\ensuremath{\mathbf p}}
\newcommand{\bQ}{\ensuremath{\mathbf Q}}
\newcommand{\br}{{\bf r}}
\newcommand{\bR}{\ensuremath{\mathbf R}}

\begin{document}
\title{Loschmidt Echo and Lyapunov Exponent in a Quantum Disordered System}
\author{ Y.~Adamov$^{1}$, I.~V.~Gornyi$^{2,1,\dagger}$, and A.~D.~Mirlin$^{1,2,\#}$}
\affiliation{$^1$Institut
f\"ur Nanotechnologie, Forschungszentrum Karlsruhe, 76021 Karlsruhe,
Germany
\\
$^2$Institut f\"ur Theorie der Kondensierten Materie, Universit\"at
Karlsruhe, 76128 Karlsruhe, Germany
}
\pacs{05.45.Mt, 03.65.Sq, 73.23.-a}
\begin{abstract}
We investigate the sensitivity of a disordered system with diffractive
scatterers to a weak external perturbation. Specifically, we calculate
the fidelity $M(t)$ (also called the Loschmidt echo) characterizing a return
probability after a propagation for a time $t$ followed by a backward
propagation governed by a slightly perturbed Hamiltonian.
For short-range scatterers we perform a diagrammatic calculation
showing that the fidelity decays first exponentially
according to the golden rule, and then follows a power law governed by the
diffusive dynamics. For long-range disorder (when the diffractive
scattering is of small-angle character) an intermediate regime emerges
where the diagrammatics is not applicable. Using the path integral
technique, we derive a kinetic equation and show that $M(t)$ 
decays exponentially with a rate governed by the classical
Lyapunov exponent.

\end{abstract}

\date{\today}

\maketitle

\section{Introduction}
\label{s1}

Quantum manifestations of the classical chaotic dynamics represent a
central issue for the field of quantum chaos. To characterize
quantitatively the stability of quantum motion, Peres
\cite{Peres84,PeresBook} proposed to consider the fidelity
\begin{equation}
  \label{eq:fidelity}
  M(t)=|\langle\psi|\exp(i\hat{H}'t)\exp(-i\hat{H}t)|\psi\rangle|^{2},
\end{equation}
where $\hat{H}'$ differs by a small perturbation from the Hamiltonian
$\hat{H}$ of
the system under consideration, and $|\psi\rangle$ is some original
state (wave packet). The quantity (\ref{eq:fidelity}) is the
probability to return into the state $|\psi\rangle$ after propagation
for a time $t$ governed by the Hamiltonian $\hat{H}$ followed by a backward
propagation with a slightly perturbed Hamiltonian $\hat{H}'$. Recently,
Jalabert and Pastawski \cite{Jalabert01} argued that for a system
whose classical counterpart is chaotic the fidelity  
(\ref{eq:fidelity}) will decay exponentially in time, with the rate
given by the classical Lyapunov exponent. Their work was motivated by
measurements of a spin-echo decoherence rate in nuclear magnetic
resonance experiments \cite{PastawskiSpinEcho}, 
and they gave a name ``quantum Loschmidt echo''
to the overlap (\ref{eq:fidelity}). The paper \cite{Jalabert01}
triggered a considerable outbreak of research activity devoted to
the sensitivity of quantum chaotic systems to external perturbations. 
In a number of subsequent publications 
\cite{Jacquod01,Cucchietti02a,Cucchietti02b,Benenti02a,
Silvestrov02a,Cerruti02,Prosen02,Silvestrov02b,Benenti02b,Jacquod02,
Zurek02,wisniacki02,prosen02b,prosen02c} the Loschmidt echo was studied
(predominantly by means of numerical 
simulations) for a variety of classically chaotic systems and its
relation to decoherence problems was discussed. These numerical works
have confirmed the key prediction of \cite{Jalabert01} that in an
appropriate parameter range the decay rate of the Loschmidt echo is
governed by the classical Lyapunov exponent. 

In the present paper, we study the Loschmidt echo (\ref{eq:fidelity})
in a different context, namely that of a {\it quantum disordered}
system. Specifically, we consider a particle moving in a weak {\it
quantum} random potential. The word ``quantum'' means here that the
scattering on this disorder is of diffractive nature. For a Gaussian
random potential assumed here this is equivalent to the condition
\be
\label{e1}
d\ll l_s,
\ee
where $l_s$ is the quantum mean free path. This situation should be
contrasted with the opposite case of a classical disorder,
for which the disorder-induced contribution to the action on a
distance $\sim d$ is much larger than $\hbar$, and the representation
of propagators in terms of a sum over classical orbits (as used
e.g. in \cite{Jalabert01}) is justified. On the other hand, the
standard theoretical tool for the quantum-disorder regime is the
impurity diagram technique. It is therefore natural to attempt to
apply the diagrammatics to the Loschmidt echo problem.

We show that indeed the diagrammatic technique can be used to
calculate the fidelity  for short times (where it is given simply by
the golden rule formula), as well as for sufficiently long times
(where it decays according to a  power law reflecting the
diffusive character of the classical motion). We demonstrate however
that for a sufficiently smooth (but still quantum as defined by
(\ref{e1})) disorder an intermediate time range emerges, where the
diagrammatic method breaks down. Using the path-integral approach, we
calculate the Loschmidt echo in this regime and find that it does show
the decay governed by the classical Lyapunov exponent, which is highly
non-trivial in view of the diffractive character of disorder. 


The rest of the paper is organized as follows. In Sec.~\ref{s2} we
consider the case of a short-range disorder ($d\ll \lambda_0$, where
$\lambda_0$ is the electron wavelength) when the diagrammatic
calculation works in the whole range of times. We identify diagrams
corresponding to the short-time (golden rule) and the long-time
(diffusion-induced power law) behavior of the fidelity and evaluate
them. Section~\ref{s3}, which is the central one for the paper, is
devoted to the case of a long-range random potential ($d\gg\lambda_0$)
when the scattering is of small-angle character. Our conclusions are
summarized in Sec.~\ref{s4}. In particular, we discuss there a 
connection between the Loschmidt echo problem and a recent activity
\cite{aleiner96,aleiner97,agam00,GM02,vavilov02} devoted to quantum
interference effects in the regime of quantum chaos.

\section {  Loschmidt Echo for the short range potential}
\label{s2}

As discussed in Sec.~\ref{s1}, we consider a model of the particle
moving in a random potential inducing a quantum (diffractive) scattering.
The Hamiltonians $H$ and $H'$ describing the forward and the backward
propagation in (\ref{eq:fidelity}) 
correspond to two slightly different potentials,
\begin{equation}
\label{eq:Hamiltonian}
\hat{H}=\frac{{\hat p}^2}{2m} + U;
\quad
\hat{H}'=\frac{{\hat p}^2}{2m} + U',
\end{equation}
where $m$ is a mass of the particle.
In this section we study the case of a short-range disorder, with the
correlation length $d\ll \lambda_0$, which is essentially equivalent
to a $\delta$-correlated (white-noise) random potential.
Thus, we have the following expressions for the correlators
\begin{equation}
\label{eq:correlator1}
\langle U(\br_1)U(\br_2) \rangle =
\langle U'(\br_1)U'(\br_2) \rangle
= \frac{1}{2\pi\nu\tau} \delta(\br_1-\br_2),
\end{equation}
where $\tau$ is the mean free time (to simplify notations, we assume
it to be exactly the same for $U$ and $U'$), and 
$\nu$ is a density of states at the Fermi energy.
The difference between the potentials $\delta U=U'-U$ is characterized
by another time scale $\tilde{\tau}$,
\begin{equation}
\label{e2}
\langle \delta U(\br_1) \delta U(\br_2) \rangle 
= \frac{1}{\pi\nu\tilde{\tau}} 
\delta(\br_1-\br_2).
\end{equation}
Clearly, we want to study the effect of a weak perturbation 
$\delta U\ll U$ or, in the other words, $\tilde{\tau}\gg\tau$. 
Finally, we take the initial state $|\psi\rangle$ in the form of a 
Gaussian wave packet 
\begin{equation}
  \label{eq:psi}
  \psi(\br)=\left(  \frac{1}{\pi\sigma^{2}}\right)  ^{D/4}%
  \exp\left[  \mathrm{i}\bp_{0}\cdot  \br%
    -\frac{\br^{2}}{2\sigma^{2}} \right], 
\end{equation}
where $\sigma\gg \lambda_0=2\pi/p_{0}$ is a width of the packet 
and $D$ is a number of space dimensions (we set $\hbar=1$ throughout
the paper).

\begin{figure}[ht]
\includegraphics{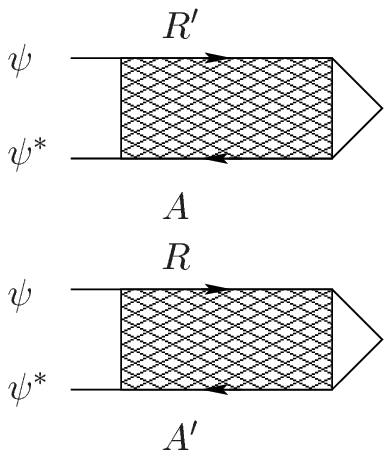}
\caption{\label{fig:diagrams} Diagram determining the short-time
(golden-rule) behavior of the fidelity.}
\end{figure}

To translate~(\ref{eq:fidelity}) into the diagrammatic language, 
we represent the ensemble-averaged fidelity as an
average product of four Green's functions
\begin{eqnarray}
  \label{eq:M(t)-2}
  &&\!\!\!\!\!\!\!\!M(t) = \int d\br_1\dots d\br_6
 \left\langle \psi(\br_1) G^{R}(\br_1,\br_2;t) G^{A'}(\br_2,\br_3;-t) 
  \right.\nonumber\\ 
  &&\!\!\!\!\!\!\!\!\left.\times \psi^{*}(\br_3)
\psi(\br_4) G^{R'}(\br_4,\br_5;t) G^{A}(\br_5,\br_6;-t)
 \psi^{*}(\br_6)\right\rangle,
\end{eqnarray}
where 
\begin{eqnarray}
  \label{eq:green}
  G^{R,A} &=& \int {dE\over 2\pi} (E-\hat{H}\pm i0)^{-1}e^{-iEt},\nonumber\\
  G^{R',A'} &=& \int {dE\over 2\pi} (E-\hat{H}'\pm i0)^{-1}e^{-iEt}.
\end{eqnarray}
The diagrams are then obtained by connecting four lines representing
the Green's functions in (\ref{eq:M(t)-2}) via the diffuson ladders.
At sufficiently short times the leading contribution is given by the
simplest diagram shown in Fig.\ref{fig:diagrams}.

\begin{figure}[ht]
\includegraphics[width=8.5cm]{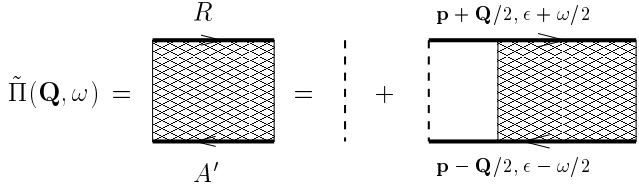}
\caption{\label{fig:diffusonshort} Dyson equation for the ``massive
diffuson''. The dashed 
  line corresponds to the correlator $\langle U U' \rangle$.}
\end{figure}

The solid lines  in Fig.\ref{fig:diagrams} correspond to the
impurity-averaged Green's functions,
\begin{eqnarray}
  \label{eq:G-R-G-A}
  \bar{G}^{R,A}(\epsilon, \bp) &=& 
  \bar{G}^{R',A'}(\epsilon, \bp) =
  \frac{1}{\epsilon-\epsilon_p\pm i/2\tau},
\end{eqnarray}
with $\epsilon_p=p^2/2m$, and  
the shaded box represents the  diffuson (see Fig.~\ref{fig:diffusonshort}), 
\begin{equation}
  \label{eq:diffusonshort}
  \tilde{\Pi}(\bQ,\omega)=\frac{1}{2\pi\nu\tau^2 
({\mathcal D}Q^{2}-i\omega+1/\tilde{\tau})},
\end{equation}
where ${\mathcal D}= v_0^2\tau/D$ is the diffusion coefficient and
$v_{0}=p_{0}/m$.  Note that this diffuson has a non-zero ``mass''
$1/\tilde{\tau}$, since it represents an averaged product of two
Green's functions in different potentials, $\langle G^RG^{A'}\rangle$.
We will only need this diffuson for zero momentum $\bQ$ and integrated
with two Green's functions, therefore it is convenient to introduce
\begin{eqnarray}
  \tilde{\Gamma}(\omega) &=& \int \frac{d\bp}{(2\pi)^D} 
  G^{R}(\epsilon+\frac{\omega}{2},\bp)G^{A'}(\epsilon-\frac{\omega}{2},\bp)
  \tilde{\Pi}({\mathbf 0},\omega)\nonumber \\
& = & \frac{1}{\tau(-i\omega+1/\tilde{\tau})}.
\end{eqnarray}
To shorten notations, we will also denote 
$\bar{G}^{R,R'}_{\psi}(\epsilon,\bp)=\psi(\bp)\bar{G}^{R,R'}(\epsilon,\bp)$
and
$\bar{G}^{A,A'}_{\psi}(\epsilon,\bp)=\psi^{*}(\bp)\bar{G}^{A,A'}(\epsilon,\bp)$,
where  $\psi(\bp)=(4\pi\sigma^{2})^{D/4}e^{-(\bp-\bp_0)^2\sigma^2/2}$ is
the wave function in the momentum representation.

With the above definitions, the expression corresponding to   
the diagram Fig.~\ref{fig:diagrams} has the form
\begin{eqnarray}
  \label{eq:exponential-pre-answer}
  M(t) &=& \left|
    \int \frac{d\bp}{(2\pi)^{D}}
    \frac{d\epsilon\;d\epsilon'}{(2\pi)^{2}}
    e^{-i(\epsilon-\epsilon')t}
  \right. \nonumber \\
  &\times& \left. \bar{G}^{R}_{\psi}(\epsilon,\bp)
    \tilde{\Gamma}(\epsilon-\epsilon')
    \bar{G}^{A'}_{\psi}(\epsilon',\bp)
  \right|^{2}.
\end{eqnarray}
After a straightforward calculation, we get the following result for
the fidelity,
\begin{eqnarray}
  \label{eq:exponential-answer}
  M(t)&=& e^{-2t/\tilde{\tau}}.
\end{eqnarray}  
This is nothing but the golden rule decay induced by the perturbation
(\ref{e2}). 

\begin{figure}[ht]
\includegraphics[width=8.5cm]{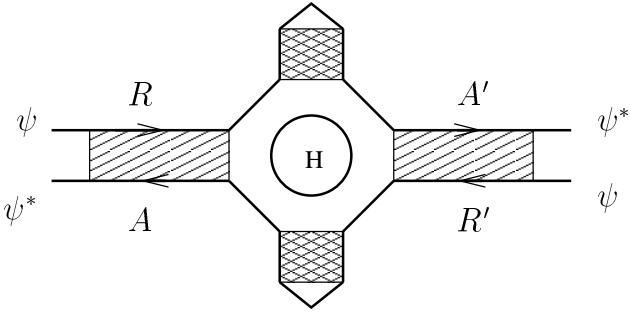}
\caption{\label{fig:hikami+diffusons}  Diagram determining the
long-time diffusive asymptotics of $M(t)$.}
\end{figure}

\begin{figure}
  \includegraphics[width=8.5cm]{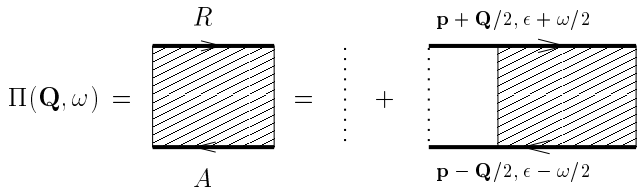}
  \caption{\label{fig:longdiffuson}
    Dyson equation for the conventional diffuson
    $\Pi(\bQ,\omega)$. The dotted  line corresponds to the
    correlator $\langle U U \rangle=\langle U' U' \rangle$.}
\end{figure}

For long times $t\gg \tilde{\tau}$ the contribution
(\ref{eq:exponential-answer}) becomes exponentially small in view of
the massive character of the diffusons (\ref{eq:diffusonshort}). 
The long-time behavior of the fidelity is however determined by a
different diagram shown in  Fig.~\ref{fig:hikami+diffusons}, with two
massive diffusons ``colliding'' and transforming into two conventional,
massless diffusons (Fig.~\ref{fig:longdiffuson}),
\begin{eqnarray}
  \label{eq:longdiffuson}
  \Pi(\bQ,\omega) &=& \frac{1}{2\pi\nu\tau^2}\frac{1}{{\mathcal D} Q^2
  -i \omega}. 
\end{eqnarray}

\begin{figure}[ht]
\includegraphics[width=8.5cm]{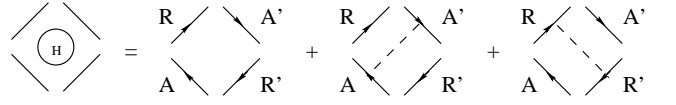}
\caption{\label{fig:Hikami-box} Diffuson ``collision'' vertex (Hikami
box).}
\end{figure}

The diffuson ``collision'' vertex, conventionally termed the Hikami
box, is given by a sum of diagrams shown in Fig.~\ref{fig:Hikami-box},
yielding 
\begin{eqnarray}
  \label{eq:Hikami}
  \chi(\bQ,\epsilon-\epsilon') &=&
  \frac{4\pi \nu\tau^2}{\tilde{\tau}}\frac{1}{(\epsilon-\epsilon')^2+(1/\tau)^2}.
\end{eqnarray}
Note the unconventional form of the expression for the Hikami box:  
due to $\langle U U'\rangle\neq \langle U U\rangle$, the diagrams
on Fig.~\ref{fig:Hikami-box} do not cancel when all diffuson momenta
and frequencies are equal to zero. 
Further, we neglected the momentum and frequencies of massless
diffusons in Eq.~(\ref{eq:Hikami}); the corresponding terms will be
smaller by a factor $\tilde{\tau}/t\ll 1$.

Combining everything, we thus get the following expression
corresponding to the diagram
Fig.~\ref{fig:hikami+diffusons}
\begin{eqnarray}
  \label{eq:crocodile}
  &&  M(t) = \int \frac{d\bp\; d\bp'\; d\bQ }{(2\pi)^{3D}}
  \int \frac{d\epsilon\; d\epsilon'\; d\omega_1\; d\omega_2}{(2\pi)^{4}}
  e^{-i(\omega_1+\omega_2)t}  \nonumber \\
&&\times  \bar{G}^{R}_{\psi}(\bp-\frac{\bQ}{2},\epsilon+\frac{\omega_{1}}{2})
  \bar{G}^{A}_{\psi}(\bp+\frac{\bQ}{2},\epsilon-\frac{\omega_{1}}{2})
\nonumber\\
&&\times  \Pi(\bQ,\omega_1)\chi(\bQ,\epsilon-\epsilon')
|\tilde{\Gamma}(\epsilon-\epsilon')|^{2}
  \Pi(\bQ,\omega_2)  \nonumber\\
&&\times  \bar{G}^{A'}_{\psi}(\bp'-\frac{\bQ}{2},\epsilon'-\frac{\omega_{2}}{2})
  \bar{G}^{R'}_{\psi}(\bp'+\frac{\bQ}{2},\epsilon'+\frac{\omega_{2}}{2}).
\end{eqnarray}
Performing all the energy integrations, we get the following result:
\begin{eqnarray}
  \label{eq:M(t)-intermediatef}
&&  M(t) = \int \frac{d\bp d\bp' d\bQ}{(2\pi)^{3D}}
  e^{-2{\mathcal D}Q^{2}t}e^{-Q^{2}\sigma^{2}/2}
  e^{-(\bp-\bp_{0})^{2}\sigma^{2}} \nonumber\\
&&\times e^{-(\bp'-\bp_{0})^{2}\sigma^{2}}
  \frac{(4\pi\sigma^{2})^{D}}{\pi\nu\tau}
  \frac{1}{(\epsilon_{\bp}-\epsilon_{\bp'})^2+\frac{1}{\tau^{2}}}.
\end{eqnarray}
Before writing down the final result, we should be more specific about
the width $\sigma$ of the original wave packet.
When it is large compared to the mean free path, 
$\sigma\gg v_{0}\tau$, the characteristic deviations
$|\bp-\bp_0|$, $|\bp'-\bp_0|$ are of the order of $1/\sigma$ due to
Gaussian factors, and we can set $|\bp|=|\bp'|$ in the last factor in 
(\ref{eq:M(t)-intermediatef}). In the opposite case $\sigma\ll
v_{0}\tau$ the difference $|\bp|-|\bp'|$ is of order $1/(v_0\tau)\ll
1/\sigma$ and  can be neglected in the above Gaussian factors. Thus we have
\begin{eqnarray}
\label{eq:answer1-large-sigma}
&&\!\!\!\!\!\!\!\!\!\!\!\!\!\!\!\!\!\!M(t) =
\frac{v_{0}\tau}{2\pi\sigma}\frac{\Gamma(D/2)}{(p_0\sigma)^{D-1}}
\left(\frac{2\sigma^2}{4{\mathcal D}t+\sigma^2}\right)^{\frac{D}{2}},\quad
 \sigma \gg v_0 \tau,\\
\label{eq:answer1-small-sigma}
&&\!\!\!\!\!\!\!\!\!\!\!\!\!\!\!\!\!\!M(t)=
 \frac{1}{2\sqrt{2\pi}}\frac{\Gamma(D/2)}{\left(p_0\sigma\right)^{D-1}}
\left(\frac{\sigma^2}{2 {\mathcal D} t}\right)^{\frac{D}{2}},\quad
\sigma \ll v_0 \tau,
\end{eqnarray}
where $\Gamma(x)$ is the Euler gamma function.
Remarkably, this large-$t$ behavior of the Loschmidt echo
is independent of $\tilde{\tau}$, i.e. of the perturbation strength.

\begin{figure}
  \includegraphics{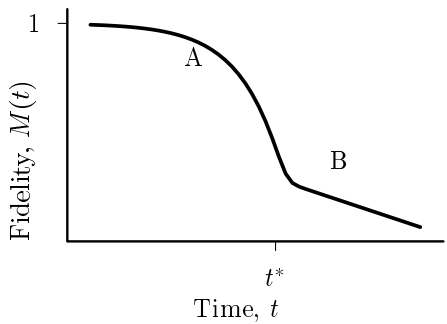}
  \caption{ \label{fig:logloggraph0}
    Schematic representation of the time evolution of 
the Loschmidt echo $M(t)$ for a white
    noise disorder on a log-log plot: A -- golden rule exponential
    decay (\ref{eq:exponential-answer}), B -- diffusive power law decay
    (\ref{eq:answer1-large-sigma}), (\ref{eq:answer1-small-sigma}).
  }
\end{figure}

Therefore, the long-time asymptotic behavior of the fidelity
is a power-law decay governed by the diffusion, with
$M(t)$ proportional to the inverse diffusion volume 
$V_{\mbox{\rm diff}}=({\mathcal D}t)^{D/2}$. 
Clearly, this result obtained from the diffuson diagram technique
is not specific for the white-noise disorder considered in this
section but rather yields a generic form of the long-$t$ 
behavior of $M(t)$ in diffusive systems.
Comparing the  results (\ref{eq:answer1-large-sigma}),
(\ref{eq:answer1-small-sigma}) with contribution  
(\ref{eq:exponential-answer}), one can determine the time $t^*$ of crossover
between the exponential and the power-law regimes, which is larger than
$\tilde{\tau}$ by a logarithmic factor. In particular, for a
two-dimensional system with $\sigma<l$ we have
$t^{*}=\frac{\tilde{\tau}}{2}\ln\frac{p_{0}{\mathcal D}\tilde{\tau}}{\sigma}$. 
The behavior of $M(t)$ for the case of a short-range potential
studied in this section is illustrated schematically 
in Fig.~\ref{fig:logloggraph0}.

It is worth mentioning that our diagrammatic calculation bears a
certain similarity to earlier studies of intensity fluctuations and
correlations for waves propagating in  random media
\cite{SpivakZyuzin37,ShapiroPRL86,Stephencollect81}.
In particular, sensitivity of transport quantities to a small change
of the impurity potential (e.g. due to a displacement of a single
scattering center) has been investigated
\cite{AltshulerSpivakJETPL85,SpivakZyuzin37}. However, the Loschmidt
echo is essentially different from the quantities studied in these
papers since it involves propagation in two different potentials $U$,
$U'$ already before the ensemble averaging. Formally, this corresponds
to switching between the Green's functions $G$ and $G'$ in external
vertices (see Figs.~\ref{fig:diagrams},\ref{fig:hikami+diffusons}).


\section {  Loschmidt Echo for long range potential}
\label{s3}

After having understood the behavior of the fidelity in a white-noise
disorder, we turn to the case of our main interest: a long-range
potential with a correlation length $d\gg\lambda_0$. 
For this kind of potential, the characteristic angle of
diffraction for each scattering event is small, $\delta \phi \sim
\lambda_0/d$, so that many scattering events are 
needed to change strongly the velocity direction.  
As a result, the motion in such a disorder is characterized by two
relaxation times. The first one is the quantum (or, in another
terminology, single-particle) relaxation time
$\tau_s$, which is  the mean time
between scattering events. This time determines the decay rate of the averaged
Green's function $\langle
G^R(\br,t)\rangle=G^R_{0}(\br,t)e^{-t/2\tau_s}$.  
The second one, the momentum (or, transport) relaxation time $\tau_{\rm tr}$
sets the time scale on which the velocity direction changes by
an angle of order $\pi$. It is parametrically larger, 
$\tau_{\rm tr} \sim (d/\lambda_0)^2 \tau_s$.  The transport time
determines, in particular,  
the diffusion coefficient ${\mathcal D}=v_0^2 \tau_{\rm tr}/D$.

The condition (\ref{e1}) implies that in the diagrammatic approach 
the leading contribution is given by diagrams with non-crossing
impurity lines. In particular, $\tau_s$ is determined by the
Born-approximation self-energy diagram, $\tau_{\rm tr}$
is obtained by taking into account the ladder-type vertex correction, and
the diffusion propagator can be calculated by solving the equation
corresponding to the sum of the ladder diagrams \cite{woelfle84}. 
Furthermore, even in the ballistic range of frequency and momenta
$ql_{\rm tr}>1$, $\omega\tau_{\rm tr}>1$, the average product of two
Green's functions $\langle G_RG_A\rangle$ is determined by a sum of
ladder diagrams, the ``ballistic diffuson''.  
One might
thus expect that the diagrammatic calculation of the preceding section
can be generalized to the case of a long-range disorder. 
Indeed, both the golden rule
short-time behavior corresponding to Fig.~\ref{fig:diagrams} 
and the long-time diffusion power-law asymptotics determined by the
diagram of Fig.~\ref{fig:hikami+diffusons} do retain their validity 
for a long-range disorder. However, as we demonstrate below, an
intermediate time range  emerges, where the diagrammatic approach is
not applicable. The reason for this is a necessity to average a
product of four Green's functions describing four electronic
trajectories propagating close to each other. As was shown in 
\cite{GM02}, in a certain time range (specified below) 
these four Green functions do not decouple into two
(ballistic) diffusons, but rather are coupled all together by impurity
correlators into a more complicated object, a ``4-diffuson''. 
In view of the failure of the ballistic-diffuson 
diagrammatics, we will use the
path integral approach developed  in \cite{MAW,GM02}. For simplicity,
we consider a two-dimensional system. 

We begin by defining 
the disorder correlation functions (replacing the white-noise
formulas (\ref{eq:correlator1}) and (\ref{e2})),
\begin{eqnarray}
\langle U(\br) U(\br_1) \rangle &=&
\langle U'(\br) U'(\br_1) \rangle=
W(|\br -\br_1|),\nonumber\\
\langle \delta U(\br) \delta U(\br_1) \rangle &=&
2\delta W(|\br -\br_1|).
\end{eqnarray}
Introducing the Feynman path integral representation and averaging
over the disorder we rewrite 
the product of four Green's functions in~(\ref{eq:M(t)-2}) as 
\begin{eqnarray}
\label{eq:GGGG}
&&\langle G_R(\bR_1,\bR_2,T)G_{A'}(\bR_2,\bR_3,-T) \nonumber \\
&& \times G_{R'}(\bR_4,\bR_5,T)
G_{A}(\bR_5,\bR_6,-T)\rangle =\nonumber\\
&&=\int_{\br_1(0)=\bR_1}^{\br_1(T)=\bR_2}
\int_{\br_3(0)=\bR_2}^{\br_3(T)=\bR_3}
\int_{\br_2(0)=\bR_4}^{\br_2(T)=\bR_5} \nonumber \\ &&
\times \int_{\br_4(0)=\bR_5}^{\br_4(T)=\bR_6}
 \prod_{i=1}^{4}{\cal
D}\br_i \quad\! \exp[iS_{\rm kin}-S_{\rm W}];
\label{eq:PI}\\
&&S_{\rm kin}={m\over 2}\int_{0}^{T}dt\left( {\dot \br}_1^2+
 {\dot \br}_2^2 - {\dot \br}_3^2- {\dot \br}_4^2\right);\nonumber \\
&&S_{\rm W}={1\over 2}(S_{11}+S_{22}+S_{33}+S_{44}) \nonumber \\
&& +S_{12}+S_{34}-S_{13}-S_{14}-S_{23}-S_{24} \nonumber \\
&& -\delta S_{12}-\delta S_{34}+\delta S_{13} 
+\delta S_{24};\nonumber \\
&&S_{ij}=\int_0^{T}\int_0^{T}
W(\br_i(t)-\br_j(t'))dt dt',\nonumber\\
&&\delta S_{ij}=\int_0^{T}\int_0^{T}
\delta W(\br_i(t)-\br_j(t'))dt dt',
\end{eqnarray}
where the paths $\br_1,\br_2$ correspond to
the retarded, and $\br_3,\br_4$ to the advanced Green's
functions. The path integral (\ref{eq:PI}) is similar to the one
evaluated in Ref.~\cite{GM02}, a difference being in boundary
conditions and in the additional terms $\delta S_{ij}$ in the action
induced by the perturbation $\delta U$.  As in \cite{GM02}, 
it is useful to perform the change of variables, introducing
$
\bR_+=(\br_1+\br_2+\br_3+\br_4)/4, \
\bR_-=(\br_1+\br_2-\br_3-\br_4), \
\br_{+}=(\br_1-\br_2+\br_3-\br_4)/2, \
\br_{-}=(\br_1-\br_2-\br_3+\br_4)/2.
$
The boundary conditions in terms of the new variables are as follows.
At $t=T$ we have
$\bR_{-}(T)=0$ and $\br_{-}(T)=0$, while 
the integration over $\bR_{+}(T)$ and
$\br_{+}(T)$ are unrestricted.
At $t=0$ the integration over $\bR_\pm(0)$ and $\br_\pm(0)$ is
performed with the weight
\begin{eqnarray}
  \label{eq:psipsipsipsi}
   &&\psi(\bR_1)\psi^{*}(\bR_3)\psi(\bR_4)\psi^{*}(\bR_6) =
   \left(\frac{1}{\pi\sigma^2}\right)^{2} \nonumber \\
&&\times   \exp\left\{-\frac{4\bR_{+}(0)^2+\bR_{-}^2(0)/4+\br_{+}(0)^2+\br_{-}(0)^2}
     {2\sigma^2} \right. \nonumber \\
&& \qquad\qquad \left. + i\bp_{0}\bR_{-}(0)\right\}.
\end{eqnarray}
The kinetic part of the action reads in the transformed variables as
\begin{equation}
  \label{eq:S-kin-rho}
  S_{\rm kin}=m\int_{0}^{T}dt\left( \dot{\br}_{+}\dot{\br}_{-} + 
    \dot{\bR}_{+}\dot{\bR}_{-}\right).
\end{equation}
As was shown in \cite{GM02}, the pairs of variables $(\bR_+,\bR_-)$ and 
$(\br_+,\br_-)$ decouple. On ballistic distances $\ll l_{\rm tr}$ 
the integral over the
first pair is essentially of the free-particle type, and its saddle-point yields
the classical equation of motion for the ``center of mass'' coordinate $\bR_+$,
\begin{equation}
  \label{eq:R(t)}
  \bR_+(t)=\bR_{+}(0) + (\bR_{+}(T)-\bR_{+}(0))\frac{t}{T}.
\end{equation}
After integrating out $\bR_+,\bR_-$, the action
(\ref{eq:S-kin-rho}) is reduced to the form \begin{eqnarray}
  \label{eq:S-kin-2}
  S_{\rm kin} &=&
  m\frac{\bR_{+}(T)-\bR_{+}(0)}{2T}(\bR_{-}(T)-\bR_{-}(0)) \nonumber \\
&+&  m\int_{0}^{T}dt\dot{\br}_{+}\dot{\br}_{-}. 
\end{eqnarray}

Since we are interested in the ballistic scales ($\ll
l_{\rm tr}$), it is convenient to split $\bR_-,\br_{+},\br_{-}$
into components parallel ($||$) and perpendicular
($\perp$) to the direction of the motion $\dot{\bR}_{+}$.
Then the disorder-induced 
part of the action $S_{\rm W}$ depends only on the transverse
components $\bR_{-\perp}$ and $\br_{\pm\perp}$, 
which we will denote $Y_{-}$ and $\rho_{\pm}$ respectively, 
\begin{eqnarray}
&& S_{\rm W} \simeq
\int_0^T\!\!{\cal U}(\rho_{-}(t),\rho_{+}(t))dt -2\int_0^T \!dt\; Y_{-}^2(t)
 \nonumber \\
&&\times\{G(\rho_{-}(t))-\delta G(\rho_{-}(t))+G(\rho_{+}(t))\},
\label{eq:SWFG}
\end{eqnarray}
where
\begin{eqnarray}
  \label{eq:U}
  &&{\cal U}={\cal U}_0+\delta {\cal U},\nonumber\\ 
  &&{\cal U}_0 =
  2(F(\rho_+)+F(\rho_-))-F(\rho_++\rho_-)-F(\rho_+-\rho_-),\nonumber\\ 
  &&\delta{\cal U} =-2\delta F(\rho_-)+\delta F(\rho_++\rho_-)+\delta
  F(\rho_+-\rho_-), 
\end{eqnarray}
and we have introduced the functions
\begin{eqnarray}
  \label{eq:F}
  &&F(y)\equiv\int_0^\infty{dx\over v_0}[W(x,0)-W(x,y)], \nonumber \\
&&  \delta F(y)\equiv\int_0^\infty{dx\over v_0}[\delta W(x,0)-\delta W(x,y)],\\
  \label{eq:defG}
  &&G(y)\equiv\int_0^\infty{dx\over v}
  {\partial^2\over \partial y^2} W(x,y), \nonumber\\
&&  \delta G(y)\equiv\int_0^\infty{dx\over v}
  {\partial^2\over \partial y^2}\delta W(x,y).
\end{eqnarray}
Since the correlation functions $W(r)$, $\delta W(r)$ decay on the
scale $d$,
the functions $F$, $G$, $\delta F$, and $\delta G$
have the following asymptotic behavior:
$F(y\ll d)\simeq -G(0)y^2/2$, $F(y\gg d)\simeq
\tau_s^{-1}$, $G(0)=-m^2v^2/\tau_{\rm tr}$, and $G(y\gg d)\to 0$,
and analogously $\delta F(y\ll d)\simeq -\delta G(0)y^2/2$, 
$\delta F(y\gg d)\simeq
\tilde{\tau}_s^{-1}$, $\delta G(0)=-m^2v^2/\tilde{\tau}_{\rm tr}$,
$\delta G(y\gg d)\to 0$.
Here the times $\tau_s$ and $\tau_{\rm tr}$ are defined according to
\begin{eqnarray}
\label{eq:tau_s}
&&\frac{1}{\tau_{s}}=\frac{2}{v_{0}}\int_{0}^{\infty}W(r)dr,
\end{eqnarray}
\begin{eqnarray}
  \label{eq:tau-tr}
  \frac{1}{\tau_{\rm tr}}=-\frac{1}{m^{2}v_{0}^{3}}
\int_{0}^{\infty}\frac{dr}{r}\frac{dW(r)}{dr}. 
\end{eqnarray}
As shown in \cite{MAW}, these are exactly the expressions for the
single-particle and the transport times in a long-range disorder.
The times $\tilde{\tau}_{s}$ and $\tilde{\tau}_{\rm tr}$ are defined by the
equations analogous to (\ref{eq:tau_s}), (\ref{eq:tau-tr}) but with a
substitution $W\to\delta W$.

Taking into account the boundary conditions~(\ref{eq:psipsipsipsi})
and integrating out the variables $\bR_{\pm}$,
we get the following expression for the fidelity
\begin{equation}
\label{eq:M(t)-integrated}
M(T)=\int\frac{d\rho_{+}d\rho_{-}}{\sqrt{2\pi\sigma^2}} 
e^{
-\frac{\rho_{-}^2+\rho_{+}^2}{2\sigma^2}
}
g(\rho_{+},\rho_{-};T)
\end{equation}
The function $g$ entering (\ref{eq:M(t)-integrated})
is determined by the $\rho_{\pm}$-part of the path
integral, which can be reduced in the standard way to a differential
equation: 
\begin{eqnarray}
\label{eq:evolutiong}
\left({\partial \over {\partial t}}
-{i\over {m}}
\frac{\partial^2}{\partial\rho_{+}\partial\rho_{-}}+
{\cal U}(\rho_{+},\rho_{-})
\right)g(\rho_{+},\rho_{-},t)&&\nonumber\\
=\delta(t)\delta(\rho_{-}).&&
\end{eqnarray}
The l.h.s. of this equation reduces to that of Eq.(36) in \cite{GM02}
if the forward and backward evolution are performed in the
same potential, $\delta{\cal U}=0$ (the r.h.s. differs from
\cite{GM02} because of different boundary conditions).
The presence of $\delta{\cal U}$ [which enters the ``potential energy''
${\cal U}$, see Eq.~(\ref{eq:U})]  
in (\ref{eq:evolutiong}) is crucially important: 
otherwise the solution would be simply
$g(\rho_+,\rho_-,t)=\delta(\rho_-)$ for any $t>0$, since
the boundary condition in (\ref{eq:evolutiong}) is
independent on $\rho_+$ and ${\cal U}_0(\rho_+,0)=0$. 
After a substitution into (\ref{eq:M(t)-integrated}), this would lead
to $M(t)=1$, which is the correct result in the absence of
perturbation. 

We have therefore reduced the problem of calculation of the Loschmidt
echo in the ballistic time range to a solution of the kinetic equation
(\ref{eq:evolutiong}). 
Let us consider the time evolution of the solution of Eq.~(\ref{eq:evolutiong}).
The initial value at $t\to 0$ is $g=\delta(\rho_-)$, and as explained
above,  the time evolution  is
initially determined by the term $\delta {\cal U}$ which induces a
$\rho_+$ dependence of the solution. As a result, 
the distribution $g$ becomes quickly suppressed at $\rho_{+}\gtrsim d$ because 
\begin{equation}
\label{eq:deltaU1}
\delta {\cal U}(\rho_+,0)\simeq  \frac{2}{\tilde{\tau}_{s}}, \qquad \rho_+\gg d.
\end{equation}
Thus  for $\rho_+\gg d$ Eq.~(\ref{eq:evolutiong}) reduces to
\begin{eqnarray}
  \label{eq:1-st-evolution}  
  &&\frac{\partial}{\partial t}g+\frac{2}{\tilde{\tau}_{s}}g=0,
\end{eqnarray}
which gives an exponential decay,
\begin{equation}
  \label{eq:g-golden-rule-decay}
  g\simeq\left\{
    \begin{array}{ll}
      \delta(\rho_-), & \rho_{+} \ll d\\
      \delta(\rho_-)e^{-2t/\tilde{\tau}_s}, & \rho_{+}\gg d
    \end{array}
  \right.
\end{equation}
Therefore, for $t\gg\tilde{\tau}_{s}$ 
the function $g$ remains essentially  non-vanishing only
in the region $\rho_+,\rho_-\ll d$. In this region we can expand 
${\cal U}$ up to the leading terms in $\rho_+$ and $\rho_-$ and
rewrite Eq.~(\ref{eq:evolutiong}) in the form 
\begin{equation}
  \label{eq:U-expanded-lyapunov}
  \left({\partial \over {\partial t}}
    -{i\over {m}}
    \frac{\partial^2}{\partial\rho_{+}\partial\rho_{-}}+
    \frac{m^2}{\tau_L^3}\rho_+^2\rho_-^2+ 
    \frac{2m^2v_0^2}{\tilde{\tau}_{\rm tr}}\rho_+^2\right)g=0.
\end{equation}
We have introduced here a new time scale  
\begin{eqnarray}
\label{eq:tau_L}
\tau_L &=& \left(\frac{3}{2m^{2}v_{0}}\int_0^\infty
 \frac{dr}{r}\frac{d}{dr}\left(\frac{1}{r}
   \frac{dW(r)}{dr}\right)\right)^{-\frac{1}{3}}\nonumber
\\ &\sim & \tau_{\rm tr}\left(\frac{d}{l_{\rm tr}}\right)^{\frac{2}{3}}.
\end{eqnarray} 
As discussed below, $\tau_L$ is equal (up to a numerical coefficient) to the
inverse Lyapunov exponent in the corresponding classical problem, and
we will call it the Lyapunov time.

At the early stage of the evolution, characteristic values of $\rho_-$
are small, and the $\rho_+^2\rho_-^2$ term in  
(\ref{eq:U-expanded-lyapunov}) is small compared to the $\rho_-^2$
term. More specifically, at $t\sim \tilde{\tau}_s$ we have 
$\rho_{-} \sim \tilde{l}_s/(p_0 d)$, so that the condition for
neglecting the quartic term in this time range is $l_L\gg \tilde{l}_s$, where
$l_L=v_0\tau_L$ is the Lyapunov length. We will  assume in the sequel  
that this condition  is fulfilled \cite{note1}.
Thus, Eq.~(\ref{eq:evolutiong}) reduces to
\begin{equation}
\label{eq:diffusion-toootr}
\left({\partial \over {\partial t}}
-{i\over {m}}
\frac{\partial^2}{\partial\rho_{+}\partial\rho_{-}}+ 
\frac{2m^2v_0^2}{\tilde{\tau}_{\rm tr}}\rho_+^2\right)g=0.
\end{equation}
Performing further a Fourier transformation $\rho_+ 
\to i (mv_0)^{-1}\partial/\partial
\phi$, $\partial/\partial \rho_+ \to imv_0 \phi$,
we cast Eq.~(\ref{eq:evolutiong}) into the following form
\begin{equation}
\label{boltzmann}
\left({\partial \over {\partial t}}
+v_0\phi
\frac{\partial}{\partial\rho_{-}}- 
\frac{2}{\tilde{\tau}_{\rm tr}}{\partial^2 \over\partial \phi^2} \right)g=0.
\end{equation}
Remarkably, Eq.~(\ref{boltzmann}) has a
meaning of the Boltzmann kinetic equation for the phase-space
distribution function describing the motion in the
transverse direction characterized by the coordinate $\rho_-$, with
$v_0\phi$ playing the role of the corresponding velocity. This clarify
the meaning of $\phi$ (and explains the notation): it is the angle the
velocity vector makes with the $||$ axis. (We remind that we are
considering ballistic time scales, so that $\phi\ll 1$)
The last term in (\ref{boltzmann})  plays the role of a collision
integral and describes a diffusion process for the velocity angle. 
Solution  of this equation is a Gaussian packet 
$$
g(\phi,\rho_-) = \frac{\sqrt{3}\tilde{\tau}}{2m v_{0}^2 t^2} 
  \exp\left\{
    \left(
      \frac{3\phi\rho_-}{v_0 t}
      -\frac{3\rho_-^2}{(v_0 t)^2}
      -\phi^2
    \right)
    \frac{\tilde{\tau}_{\rm tr}}{2t}
  \right\}.
$$
 Transforming back to the variable $\rho_+$, we get
$$
  g(\rho_{+},\rho_{-}) = 
  \frac{1}{\sqrt{\pi}\Sigma_{-}}
  \exp\left\{
    -\frac{\rho_{-}^{2}}{\Sigma_{-}^{2}}
    -\frac{\rho_{+}^{2}}{\Sigma_{+}^{2}}
    +2i\sqrt{3}\frac{\rho_{+}\rho_{-}}{\Sigma_{+}\Sigma_{-}}
  \right\},
$$
where
\begin{eqnarray}
  \label{eq:sigmapm-pre-lyapunov}
  \Sigma_-(t) &=& \frac{2 v_{0}t}{\sqrt{3}} 
  \left(
    \frac{2t}{\tilde{\tau}_{\rm tr}}
  \right)^{1/2},
  \nonumber\\
  \Sigma_+(t) &=&  \frac{2}{mv_{0}} \left(
    \frac{\tilde{\tau}_{\rm tr}}{2t}
  \right)^{1/2}.
\end{eqnarray}
For the phase-space distribution function $g(\phi,\rho_-,)$ the 
quantities $\Sigma_-$ and $\Sigma_+^{-1}$ play the role of widths
of the distribution with respect to the coordinate $\rho_-$ and the
momentum $mv_0\phi$, respectively.

To simplify calculations, we will 
neglect the cross-correlations between $\rho_{+}$ and
$\rho_{-}$ (this will only influence a numerical prefactor in $M(t)$,
which is of minor importance here) and assume that $g$ has the form
\begin{eqnarray}
  \label{eq:g-sigmapm}
  g(\rho_+,\rho_-) &=& \frac{1}{\sqrt{\pi}\Sigma_{-}}
  \exp\left\{
    -\frac{\rho_{-}^2}{\Sigma_{-}^2}-
    \frac{\rho_{+}^2}{\Sigma_{+}^2}
  \right\}.
\end{eqnarray}
We will see that this form of $g$ will preserve during the further evolution
of the distribution.

The characteristic values of $\rho_-$ are
increasing proportionally to $t^{3/2}$. Eventually, the neglected
third (quartic) term in Eq.~(\ref{eq:U-expanded-lyapunov}) becomes comparable
to the fourth one. Using the result
(\ref{eq:sigmapm-pre-lyapunov}) for the characteristic value
$\Sigma_-(t)$ of $\rho_-$, it is easy to see that
this happens at $t=\tau_{L}$.
At $t>\tau_L$ the fourth term dominates, and 
equation (\ref{eq:U-expanded-lyapunov})  takes the form
\begin{equation}
  \label{eq:Evolution-Lyapunov}
  \left({\partial \over {\partial t}}
    -{i\over {m}}
    \frac{\partial^2}{\partial\rho_{+}\partial\rho_{-}}+
    \frac{m^2}{\tau_L^3}\rho_+^2\rho_-^2\right)g=0.
\end{equation}  
After the Fourier transformation from $\rho_+$ to $\phi$, the last
term takes the form $-(m^2/\tau_L^3)\partial^2g/\partial\phi^2$ and
describes a  diffusion process for the angle $\phi$ with the diffusion
coefficient proportional to to the coordinate $\rho_-$. This is
exactly the kinetic equation for the disorder-averaged distribution function
$g(\phi,\rho_-)$ of phase-space separations between two classical
paths \cite{aleiner96,GM02}.
It leads  to an exponential increase
of the width of the distribution function $g(\rho_-,\phi)$ with a rate
given by the Lyapunov exponent $\sim \tau_L^{-1}$. 
Thus in this Lyapunov regime we have
\begin{eqnarray}
  \label{eq:sigmapmlyapunov}
  \Sigma_{-}(t) &\simeq& v_{0}\tau_L 
  \left(
    \frac{\tau_L}{\tilde{\tau}_{\rm tr}}
  \right)^{1/2}
  \exp\left\{c \frac{t}{\tau_L}\right\}, \nonumber\\
  \Sigma_+(t) &\simeq& \lambda_0\left(
    \frac{\tilde{\tau}_{\rm tr}}{\tau_L}
  \right)^{1/2}\exp\left\{-c \frac{t}{\tau_L}\right\},
\end{eqnarray}   
where $c$ is a numerical coefficient of order unity.

Let us emphasize a highly non-trivial character of the emergence of a
classical kinetic equation. Indeed, we consider a random potential for
which each scattering act is of a quantum (diffractive) nature and can
not be described classically. It is only after the disorder averaging
that the classical kinetics is restored.

The Lyapunov regime breaks down when the width $\Sigma_-$ reaches the
correlation length $d$, i.e.
at $\tau_E^{*}=(\tau_L/c)\ln (\tilde{\tau}_{\rm tr}/\tau_{\rm tr})$. This
time plays a
role analogous to the Ehrenfest time but only depends on classical 
parameters (if one assumes that the perturbation $\delta U$ is
independent of $\hbar$). 

When $t>\tau_E^{*}$, so that $\rho_-\gg d$, we can use another asymptote of
the ``potential''  
${\cal U}(\rho_+,\rho_-)|_{\rho_{-}\gg d}=2m^2v_{0}^2\rho_{+}^2/\tau_{\rm tr}$.
This leads to an equation very similar to 
Eq.~(\ref{eq:diffusion-toootr}) but with $\tilde{\tau}_{\rm tr}$ replaced by
$\tau_{\rm tr}$. 
Therefore, in analogy with (\ref{eq:sigmapm-pre-lyapunov}),
we have again a power-law dependence of the distribution
widths,
\begin{eqnarray}
  \label{eq:sigmapm-ballistic}
  \Sigma_-(t) &\simeq& v_{0}t 
  \left(
    \frac{t}{\tau_{\rm tr}}
  \right)^{1/2},
  \nonumber\\
  \Sigma_+(t) &\simeq& \lambda_0\left(
    \frac{\tau_{\rm tr}}{t}
  \right)^{1/2}.
\end{eqnarray}
For times larger than $t_E^*$ all the calculations can also be performed
using the diagrammatic technique for ballistic systems.
 This is because four trajectories, which were coupled all
together into a ``4-diffuson'' by disorder correlations in the Lyapunov
regime, split now in two conventional ballistic-diffusons separated by a
distance $\gg d$ \cite{aleiner96,GM02}. 

\begin{figure}
\includegraphics[width=8.5cm]{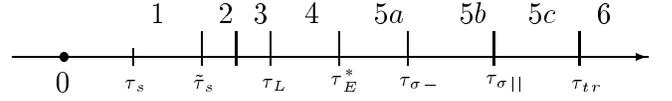}
\caption{\label{fig:line}
Characteristic time scales separating various regimes of the behavior
of $M(t)$. The regimes $2$ (golden rule) and $4$ (Lyapunov)
correspond to an exponential decay of $M(t)$, the remaining
regimes to a power-law decay (see text for details).
}
\end{figure}

Having got the solution $g$ of the kinetic equation in all the regimes
of interest, we can calculate the fidelity
$M(t)$. Substituting (\ref{eq:g-sigmapm})
into Eq.~(\ref{eq:M(t)-integrated}), we get
\begin{eqnarray} 
  \label{eq:super-final-answer}
  M(T) &\simeq& \frac{\mbox{\rm min}(\Sigma_{+}(T),\sigma)}
  {\mbox{\rm max}(\Sigma_{-}(T),\sigma)}.
\end{eqnarray}
Thus, two additional time scales may become significant
when we consider the fidelity: $\tau_{\sigma+}$, which is defined by
$\Sigma_{+}(\tau_{\sigma+})=\sigma$,   and   $\tau_{\sigma-}$ defined by
$\Sigma_{-}(\tau_{\sigma-})=\sigma$. Position of these time scales
with respect to main characteristic times ($\tau_L$, $\tau_E^*$,
$\tau_{\rm tr}$) depends on the width $\sigma$ of the initial state. 
As an example, we choose $d<\sigma<l_{\rm tr}$. In this case $\Sigma_{+}<\sigma$
for all times $t>\tilde{\tau}_s$, so that the scale $\tau_{\sigma+}$
does not arise, and $\tau_{E}^{*}<\tau_{\sigma-}<\tau_{\rm tr}$. 
The order of all characteristic scales on the time axis is illustrated
in Fig.~\ref{fig:line}.

We mention for completeness that there is one more characteristic
time $\tau_{\sigma||}$ located between
$\tau_{\sigma-}$ and $\tau_{\rm tr}$.  
At this time, the approximation~(\ref{eq:R(t)}) of straight motion 
in the coordinate $\bR_+$ loses its validity. This leads to an
additional factor 
$\sigma/\max\{\sigma,\Sigma_{||}\}$ in the expression for $M(t)$, where
\begin{eqnarray}
  \label{eq:Sigmapar}
  \Sigma_{||} &\simeq& vt \frac{t}{\tau_{\rm tr}}
\end{eqnarray}
characterizes longitudinal fluctuations of $\bR_{+}$.
There is thus an additional crossover inside the
ballistic-diagrammatics
regime, which takes place  at a time
$\tau_{\sigma||}$ satisfying $\Sigma_{||}(\tau_{\sigma||})=\sigma$.

\unitlength=2mm

We are now prepared to summarize the results of this section and to
give a list of all the regimes of behavior of the fidelity. We have
found as much as 6 essentially different regimes  (as illustrated in
Fig.~\ref{fig:line}), one of them (the 
ballistic diffusion) splits up into three subregimes with
different power-law behavior. We list the regimes in the order they
appear as the time increases:

\begin{enumerate}

\item Perfect echo regime, $t\ll\tilde{\tau}_s$. 

At such short times
the perturbation is essentially irrelevant, and $M(t)\simeq 1$.

\item Golden rule regime, $t<\tilde{\tau}_{s}\ln(\sigma/d)$.

  Substituting Eq.~(\ref{eq:g-golden-rule-decay}) into
  (\ref{eq:M(t)-integrated}) we get the exponential decay,
  \begin{equation}
    M(t) = e^{-2t/\tilde{\tau}_{s}}.
  \end{equation}

\item Power-law ``pre-Lyapunov  inflation'' regime,
 $\tilde{\tau}_{s}\ln(\sigma/d)<t<\tau_L$.

Substituting Eq.~(\ref{eq:sigmapm-pre-lyapunov}) into
  (\ref{eq:super-final-answer}) we get 
  \begin{equation}
    M(t) \sim \frac{d}{\sigma}\left(\frac{\tilde{\tau}_{s}}{t}\right)^{1/2}.
  \end{equation}
This behavior of the fidelity is related to a power-law spreading of
classical trajectories in this regime due to the ``ballistic
diffusion'' in the perturbation potential $\delta U$.

\item Lyapunov regime, $\tau_L<t<\tau_E^{*}$.
  
  Combining (\ref{eq:sigmapmlyapunov}) and (\ref{eq:super-final-answer}),
  we get an exponential decay of the Loschmidt echo determined by the
  classical Lyapunov exponent,
  \begin{equation}
    M(t) \sim \frac{d}{\sigma}
    \left(
      \frac{\tilde{\tau}_{s}}
      {\tau_L}
    \right)^{1/2}e^{-c t/\tau_L}.
  \end{equation} 

\item  ``Ballistic diffusion'' regime, $\tau_E^{*}<t<\tau_{\rm tr}$.

   This regime characterized by a power-law behavior of the fidelity
is further subdivided into three sub-regimes:
  \begin{enumerate}
  \item $\tau_E^{*}<t<\tau_{\sigma-}$.
    Using (\ref{eq:super-final-answer}), (\ref{eq:sigmapm-ballistic}) and
    noticing that $\Sigma_{-}$ is still smaller than $\sigma$ in this
time range,  we find
    \begin{equation}
      M(t) \sim \frac{1}{p_{0}\sigma}\left(\frac{\tau_{\rm tr}}{t}\right)^{1/2}.
    \end{equation}
  \item $\tau_{\sigma-}<t<\tau_{\sigma||}$.
    The only difference compared to the previous case is that now
$\Sigma_{-}>\sigma$, yielding
    \begin{equation}
      M(t) \sim \frac{1}{p_{0}l_{\rm tr}}\left(\frac{\tau_{\rm tr}}{t}\right)^{2}.
    \end{equation}
  \item $\tau_{\sigma||}<t<\tau_{\rm tr}$.
    In this regime fluctuations in the motion in the parallel
direction also become  important, see the text around
Eq.~(\ref{eq:Sigmapar}), with the result
   \begin{eqnarray}
      \label{eq:M(t)-simplebal-diffusion}
      M(t) &\sim & \frac{\sigma}{p_{0}l_{\rm tr}^2}
      \left(\frac{\tau_{\rm tr}}{t}\right)^{4}.
    \end{eqnarray} 
  \end{enumerate}
\item Conventional diffusion regime, $t>\tau_{\rm tr}$.

In this regime the nature of disorder (short-range vs. long-range) is
irrelevant, and the result (\ref{eq:answer1-small-sigma}) derived in
Sec.~\ref{s2} is applicable,
    \begin{equation}
      M(t) \sim \frac{\sigma}{p_{0}l_{\rm tr}^2}\frac{\tau_{\rm tr}}{t}.
    \end{equation}
\end{enumerate}

We would like to remind that the ordering of relevant time
scales in Fig.~\ref{fig:line} depends on the microscopic parameters
of the problem. We have considered the most interesting case
($\tilde{l}_s\ll l_L$ and $d<\sigma<l_{\rm tr}$), when all possible
regimes are developed. For other choices of parameters, some of the
regimes may disappear (see, in particular, the footnote
\cite{note1}).

\section{Conclusions} 
\label{s4}

In this paper, we have studied the Loschmidt echo (or, in a different
terminology, the fidelity), which 
characterizes the sensitivity of a quantum system to an external
perturbation, in a model with a weak quantum random potential. 
Using the diagrammatic approach, we have shown that at short times the
fidelity decays exponentially with the rate $2/\tilde{\tau}_s$ set by
the golden rule, while its long-time asymptotics is of power-law type
and is determined by the diffusive nature of the dynamics on this time
scale. For a sufficiently long-range disorder a time range emerges
where the diagrammatics becomes inapplicable due to merging of two
ballistic diffusons into a more complicated ``4-diffuson''. To study
the fidelity in this regime, we have applied a quasiclassical
(path integral) approach. This allowed us to express the
disorder-averaged fidelity in
terms of a solution of a quasiclassical evolution equation,
see Eqs.~(\ref{eq:M(t)-integrated}) and (\ref{eq:evolutiong}). On time
scales larger than $\tilde{\tau}_s$ this equation takes a form of the
kinetic equation for the distribution function $g(\phi_,\rho_-)$ of
phase-space separations between two {\it classical} paths (one of
which is subject to the perturbation). Solving the kinetic equation,
we find several additional regimes of behavior of the Loschmidt echo,
separating the short-time golden-rule decay from the long-time
diffusive asymptotics. In particular, there arises a ``Lyapunov
regime'', where the fidelity decays exponentially with a rate governed
by the classical Lyapunov exponent. 

It is worth mentioning that our path-integral calculation of the
fidelity is closely connected to the analysis of quantum interference
effects in a long-range disorder performed in \cite{aleiner96,GM02}. 
In particular, after the Fourier transformation $\rho_+\to\phi$
our evolution equation (\ref{eq:U-expanded-lyapunov})
has the same form as the equation describing the ``Hikami box'' in
\cite{aleiner96} (after averaging over the smooth random potential). 
This is a remarkable agreement, since the methods used are essentially
different: Aleiner and Larkin \cite{aleiner96} work in a given
realization of a random potential (which assumes that the potential is
classical, i.e. the condition opposite to Eq.~(\ref{e1}) is
fulfilled), while we consider the case of a diffractive scattering
[Eq.~(\ref{e1})] and perform all calculations for disorder-averaged
quantities. There is, however, an important difference between the
equations obtained. Specifically, in our case the last term of
Eq.~(\ref{eq:U-expanded-lyapunov}) (which is proportional to
$1/\tilde{\tau}_{\rm tr}$) is due to the difference $\delta U$
between the Hamiltonians for the forward and the backward
propagation. On the other hand, the authors of Ref.~\cite{aleiner96}
add ``by hand'' a term of exactly the same type (with a certain time
$\tau_q$ replacing our $\tilde{\tau}_{\rm tr}$) for a problem without
any perturbation $\delta U$, arguing that it mimics
a small-angle diffraction in the system. To our opinion, this
justification is questionable (at least, for a system with a weak smooth
disorder). Indeed, in this case all scattering processes determining
the transport in the system are of diffractive type and are taken into
account in our approach. There is thus no freedom to add an additional
``diffractive'' term to the kinetic equation. We thus believe that the
Hikami box is described by an equation without such term (i.e. analogous to
our equations (\ref{eq:evolutiong}), (\ref{eq:U-expanded-lyapunov})
in the absence of perturbation,  $\delta{\cal U}, \tilde{\tau}_{\rm
tr}^{-1}=0$) but with appropriate boundary conditions. While this will
probably not affect the main results of Ref.~\cite{aleiner96}
(depending only logarithmically on $\tau_q$), such a more consistent
treatment of the quasiclassical Hikami box \cite{note2} would be of conceptual
importance for the theory of quantum interference effects in 
systems with large-scale inhomogeneities. We leave this issue as an
open problem for the future research. 

\section{Acknowledgments}

Discussions with B. Shapiro are gratefully acknowledged. This work was
supported by the SFB195 der Deutschen Forschungsgemeinshaft, by the
German-Israeli Foundation and by the RFBR.

A part of this work was done when two of us (I.V.G. and A.D.M.) participated
in the program ``Chaos and Interactions: from Nuclei to Quantum Dots''
of the Institute for Nuclear Theory at the University of Washington,
Seattle. We thank the Institute for Nuclear Theory for hospitality and
partial support.


\begin{thebibliography}{10}
\bibitem{note1} In the opposite case $l_L\ll
\tilde{l}_s$ (which is the case if the perturbation is very weak, or
if the potential is not sufficiently long-ranged), the Lyapunov regime
does not exist, and the golden rule decay crosses over directly to
a power-law ballistic-diffusion regime.
\end{thebibliography}
\end{document}